\documentclass[
reprint,
superscriptaddress,
groupedaddress,
longbibliography,
amsmath,amssymb,
aps,
prapplied
]{revtex4-2}

\usepackage{graphicx}
\usepackage{dcolumn}
\usepackage{bm}

\usepackage[english]{babel}

\usepackage{amsmath}

\usepackage{braket}

\usepackage{hyperref}
\hypersetup{colorlinks,allcolors=blue}
\usepackage{tikz}
\usetikzlibrary{calc}
\usetikzlibrary{positioning}

\usepackage{pgfplots}
\usepgfplotslibrary{colorbrewer}
\usepgfplotslibrary{colormaps}
\pgfplotsset{width=0.48\textwidth,compat=1.9,
/pgfplots/colormap={inferno256}{
rgb255=(0.36, 0.12, 3.54)
rgb255=(1.16, 0.86, 7.83)
rgb255=(2.46, 1.97, 14.08)
rgb255=(4.27, 3.20, 21.48)
rgb255=(6.53, 4.38, 29.33)
rgb255=(9.15, 5.34, 37.23)
rgb255=(12.05, 5.96, 44.83)
rgb255=(15.11, 6.13, 51.84)
rgb255=(18.20, 5.80, 58.09)
rgb255=(21.20, 5.05, 63.40)
rgb255=(24.80, 3.60, 68.74)
rgb255=(27.74, 1.97, 71.76)
rgb255=(30.28, 0.32, 72.06)
rgb255=(32.40, 0.00, 69.72)
rgb255=(34.10, 0.00, 66.25)
rgb255=(35.47, 0.00, 62.03)
rgb255=(36.56, 0.00, 57.28)
rgb255=(37.42, 0.00, 52.17)
rgb255=(38.11, 0.00, 46.77)
rgb255=(38.59, 0.00, 41.11)
},
/pgfplots/colormap={turbo256}{%
rgb255=(48,18,59) rgb255=(50,21,67) rgb255=(51,24,74) rgb255=(52,27,81) rgb255=(53,30,88) rgb255=(54,33,95) rgb255=(55,36,102) rgb255=(56,39,109) rgb255=(57,42,115) rgb255=(58,45,121) rgb255=(59,47,128) rgb255=(60,50,134) rgb255=(61,53,139) rgb255=(62,56,145) rgb255=(63,59,151) rgb255=(63,62,156) rgb255=(64,64,162) rgb255=(65,67,167) rgb255=(65,70,172) rgb255=(66,73,177) rgb255=(66,75,181) rgb255=(67,78,186) rgb255=(68,81,191) rgb255=(68,84,195) rgb255=(68,86,199) rgb255=(69,89,203) rgb255=(69,92,207) rgb255=(69,94,211) rgb255=(70,97,214) rgb255=(70,100,218) rgb255=(70,102,221) rgb255=(70,105,224) rgb255=(70,107,227) rgb255=(71,110,230) rgb255=(71,113,233) rgb255=(71,115,235) rgb255=(71,118,238) rgb255=(71,120,240) rgb255=(71,123,242) rgb255=(70,125,244) rgb255=(70,128,246) rgb255=(70,130,248) rgb255=(70,133,250) rgb255=(70,135,251) rgb255=(69,138,252) rgb255=(69,140,253) rgb255=(68,143,254) rgb255=(67,145,254) rgb255=(66,148,255) rgb255=(65,150,255) rgb255=(64,153,255) rgb255=(62,155,254) rgb255=(61,158,254) rgb255=(59,160,253) rgb255=(58,163,252) rgb255=(56,165,251) rgb255=(55,168,250) rgb255=(53,171,248) rgb255=(51,173,247) rgb255=(49,175,245) rgb255=(47,178,244) rgb255=(46,180,242) rgb255=(44,183,240) rgb255=(42,185,238) rgb255=(40,188,235) rgb255=(39,190,233) rgb255=(37,192,231) rgb255=(35,195,228) rgb255=(34,197,226) rgb255=(32,199,223) rgb255=(31,201,221) rgb255=(30,203,218) rgb255=(28,205,216) rgb255=(27,208,213) rgb255=(26,210,210) rgb255=(26,212,208) rgb255=(25,213,205) rgb255=(24,215,202) rgb255=(24,217,200) rgb255=(24,219,197) rgb255=(24,221,194) rgb255=(24,222,192) rgb255=(24,224,189) rgb255=(25,226,187) rgb255=(25,227,185) rgb255=(26,228,182) rgb255=(28,230,180) rgb255=(29,231,178) rgb255=(31,233,175) rgb255=(32,234,172) rgb255=(34,235,170) rgb255=(37,236,167) rgb255=(39,238,164) rgb255=(42,239,161) rgb255=(44,240,158) rgb255=(47,241,155) rgb255=(50,242,152) rgb255=(53,243,148) rgb255=(56,244,145) rgb255=(60,245,142) rgb255=(63,246,138) rgb255=(67,247,135) rgb255=(70,248,132) rgb255=(74,248,128) rgb255=(78,249,125) rgb255=(82,250,122) rgb255=(85,250,118) rgb255=(89,251,115) rgb255=(93,252,111) rgb255=(97,252,108) rgb255=(101,253,105) rgb255=(105,253,102) rgb255=(109,254,98) rgb255=(113,254,95) rgb255=(117,254,92) rgb255=(121,254,89) rgb255=(125,255,86) rgb255=(128,255,83) rgb255=(132,255,81) rgb255=(136,255,78) rgb255=(139,255,75) rgb255=(143,255,73) rgb255=(146,255,71) rgb255=(150,254,68) rgb255=(153,254,66) rgb255=(156,254,64) rgb255=(159,253,63) rgb255=(161,253,61) rgb255=(164,252,60) rgb255=(167,252,58) rgb255=(169,251,57) rgb255=(172,251,56) rgb255=(175,250,55) rgb255=(177,249,54) rgb255=(180,248,54) rgb255=(183,247,53) rgb255=(185,246,53) rgb255=(188,245,52) rgb255=(190,244,52) rgb255=(193,243,52) rgb255=(195,241,52) rgb255=(198,240,52) rgb255=(200,239,52) rgb255=(203,237,52) rgb255=(205,236,52) rgb255=(208,234,52) rgb255=(210,233,53) rgb255=(212,231,53) rgb255=(215,229,53) rgb255=(217,228,54) rgb255=(219,226,54) rgb255=(221,224,55) rgb255=(223,223,55) rgb255=(225,221,55) rgb255=(227,219,56) rgb255=(229,217,56) rgb255=(231,215,57) rgb255=(233,213,57) rgb255=(235,211,57) rgb255=(236,209,58) rgb255=(238,207,58) rgb255=(239,205,58) rgb255=(241,203,58) rgb255=(242,201,58) rgb255=(244,199,58) rgb255=(245,197,58) rgb255=(246,195,58) rgb255=(247,193,58) rgb255=(248,190,57) rgb255=(249,188,57) rgb255=(250,186,57) rgb255=(251,184,56) rgb255=(251,182,55) rgb255=(252,179,54) rgb255=(252,177,54) rgb255=(253,174,53) rgb255=(253,172,52) rgb255=(254,169,51) rgb255=(254,167,50) rgb255=(254,164,49) rgb255=(254,161,48) rgb255=(254,158,47) rgb255=(254,155,45) rgb255=(254,153,44) rgb255=(254,150,43) rgb255=(254,147,42) rgb255=(254,144,41) rgb255=(253,141,39) rgb255=(253,138,38) rgb255=(252,135,37) rgb255=(252,132,35) rgb255=(251,129,34) rgb255=(251,126,33) rgb255=(250,123,31) rgb255=(249,120,30) rgb255=(249,117,29) rgb255=(248,114,28) rgb255=(247,111,26) rgb255=(246,108,25) rgb255=(245,105,24) rgb255=(244,102,23) rgb255=(243,99,21) rgb255=(242,96,20) rgb255=(241,93,19) rgb255=(240,91,18) rgb255=(239,88,17) rgb255=(237,85,16) rgb255=(236,83,15) rgb255=(235,80,14) rgb255=(234,78,13) rgb255=(232,75,12) rgb255=(231,73,12) rgb255=(229,71,11) rgb255=(228,69,10) rgb255=(226,67,10) rgb255=(225,65,9) rgb255=(223,63,8) rgb255=(221,61,8) rgb255=(220,59,7) rgb255=(218,57,7) rgb255=(216,55,6) rgb255=(214,53,6) rgb255=(212,51,5) rgb255=(210,49,5) rgb255=(208,47,5) rgb255=(206,45,4) rgb255=(204,43,4) rgb255=(202,42,4) rgb255=(200,40,3) rgb255=(197,38,3) rgb255=(195,37,3) rgb255=(193,35,2) rgb255=(190,33,2) rgb255=(188,32,2) rgb255=(185,30,2) rgb255=(183,29,2) rgb255=(180,27,1) rgb255=(178,26,1) rgb255=(175,24,1) rgb255=(172,23,1) rgb255=(169,22,1) rgb255=(167,20,1) rgb255=(164,19,1) rgb255=(161,18,1) rgb255=(158,16,1) rgb255=(155,15,1) rgb255=(152,14,1) rgb255=(149,13,1) rgb255=(146,11,1) rgb255=(142,10,1) rgb255=(139,9,2) rgb255=(136,8,2) rgb255=(133,7,2) rgb255=(129,6,2) rgb255=(126,5,2) rgb255=(122,4,3)} %
}

\pgfmathsetlengthmacro\MajorTickLength{
  \pgfkeysvalueof{/pgfplots/major tick length} * 0.7}
  
\begin{document}


    \title{Efficient site-resolved imaging and spin-state detection in dynamic\\ two-dimensional ion crystals}


\author{Robert N. Wolf}
\email[e-mail: ]{robert.wolf@sydney.edu.au}
\author{Joseph H. Pham}
\author{Julian Y. Z. Jee}%

\author{Alexander Rischka}
\altaffiliation[Current address: ]{Q-CTRL Pty Ltd, Sydney, NSW 2000, Australia}

\author{Michael J. Biercuk}%
\affiliation{%
ARC Centre for Engineered Quantum Systems, School of Physics, The University of Sydney, Sydney, NSW 2006, Australia
}%

\date{\today}

\begin{abstract}
Resolving the locations and discriminating the spin states of individual trapped ions with high fidelity is critical for a large class of applications in quantum computing, simulation, and sensing. We report on a method for high-fidelity state discrimination in large two-dimensional (2D) crystals with over 100 trapped ions in a single trapping region, combining a hardware detector and an artificial neural network. A high-data-rate, spatially resolving, single-photon sensitive timestamping detector performs efficient single-shot detection of 2D crystals in a Penning trap, exhibiting rotation at about $25\,\mathrm{kHz}$. We then train an artificial neural network to process the fluorescence photon data in the rest frame of the rotating crystal in order to identify ion locations with a success rate of $~90\%$, accounting for substantial illumination inhomogeneity across the crystal. Finally, employing a time-binned state detection method, we arrive at an average spin-state detection fidelity of $94(2)\%$. This technique can be used to analyze spatial and temporal correlations in arrays of hundreds of trapped-ion qubits.
\end{abstract}

\maketitle

\section{INTRODUCTION}

Trapped atomic ions are a well-established platform with tremendous potential for quantum computation \cite{Alexeev2021}, simulation \cite{Britton2012, Gaerttner2017, Safavi-Naini2018, Monroe2021, MacDonell2021}, and sensing \cite{Biercuk2010, Gilmore2021a, Degen2017, Marciniak2022}. One of the primary aims in all fields is to increase the number of participating ions in an experiment, either to expand the quantum information capacity of a quantum computer or to augment the performance of a quantum sensor. Paul traps have shown control of over 50 ions in one-dimensional strings \cite{Kranzl2022, Zhang2017, Li2023} and recently, two-dimensional (2D) arrays have been realized in a trapping configuration which mitigates the detrimental effect of radio-frequency (rf) induced micromotion \cite{Kiesenhofer2023, Qiao2022, Kato2022, DOnofrio2021}. Penning traps offer an alternative platform that avoids rf-induced micromotion and therefore enables the creation of stable 2D ion crystals with hundreds of participating ions \cite{Britton2012}. 

One of the major challenges in all ion traps and across all applications is the need to read out the internal state of individual ions across large spatial arrays. State detection can be performed with a form of spatially resolving detector. However, it faces complications due to physical and hardware-derived phenomena, including nonuniform illumination, spatial inhomogeneity in imaging due to aberrations, nonuniform ion spacing in a quadratic potential well, array dynamics, and instabilities. These challenges are exacerbated in Penning traps due to the fast rotation of the ion crystal induced by the $\Vec{E}\times\Vec{B}$ drift generated by the trapping fields. The extensive detection times required in spatially resolving imaging methods~\cite{Huang1998a, Mitchell1998, Mitchell2001, Ball2019, Britton2012, Jordan2019} are incompatible with single-shot discrimination. 

Here we present an automated method to read out the states of large arrays of ions in the most demanding circumstances identified above, using a high-data-rate, spatially resolving, single-photon sensitive timestamping detector augmented by machine-learning data analysis techniques. We present experimental demonstrations of automated state discrimination in the challenging setting of a fast-rotating 2D $^9\mathrm{Be}^+$-ion crystal in a Penning trap. Timing information obtained for each photon-detection event is used to transform the fluorescence photon positions recorded in the laboratory frame to the rest frame of the dynamic ion crystal (here due to the crystal rotation). A convolutional neural network localizes the individual ion positions, accounting for nonuniform ion spacing and the possible inclusion of dark ions. We then apply a time-binned maximum likelihood method in order to perform state discrimination for each ion~\cite{Woelk2015}, including repumping dynamics that occur on timescales comparable to the readout. As a result, we achieve an average state-detection fidelity of $94(2)\%$ for a crystal size of $118\pm3$ ions using a detection time of only about $10\,\mathrm{ms}$.


\begin{figure*}[h!tb]
    \centering
    \includegraphics[width=0.95\textwidth]{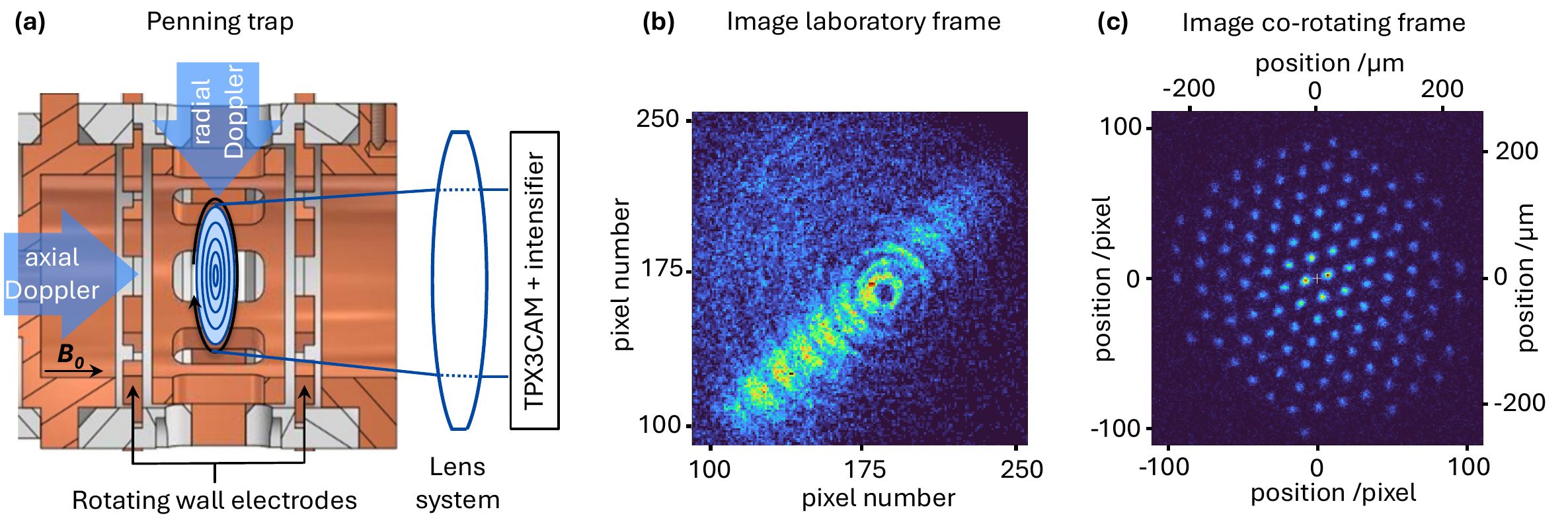}
    \caption{(a) Schematic illustration of the Penning trap with rotating 2D ion crystal, axial Doppler (waist radius $375\,\mathrm{\mu m}$, power $100\,\mathrm{\mu W}$) and radial Doppler (waist radius $45\,\mathrm{\mu m}$, power $10\,\mathrm{\mu W}$) cooling lasers, top-view lens system, TPX3CAM detection system. (b) Top-view image in the laboratory frame. (c) Derotated image of a 2D ion crystal with 121 ions.}
    \label{fig:trap}
\end{figure*}

\section{EXPERIMENTAL PLATFORM}
The Penning trap presents a challenging canonical example of an ion-state-detection problem in a multidimensional ion crystal. Here ions are nonstationary in the lab frame due to crystal rotation, making the effects encountered in this platform an extreme representation of other ion-trap experiments in which ions may suffer large micromotion at large offsets from the trap center.

Beryllium-9 ions are confined and Doppler laser cooled in a high-optical-access Penning trap system at a magnetic field strength of $B_0 \approx 2\,\mathrm{T}$ \cite{Ball2019}. Ions in a Penning trap experience $\Vec{E}\times\Vec{B}$ drift about the electrostatic center of the trap. For an ensemble of ions, the associated rotation frequency, $\omega_r/2\pi$, can be controlled via external electric fields, known as a ``rotating wall''. This is a rotating azimuthal quadrupole field generated by a pair of eight-fold segmented electrodes which phase-locks $\omega_r/2\pi$ at a value bounded by the single-particle magnetron and reduced cyclotron frequency. Two-dimensional ion crystals can be created near these two boundaries, while for practical reasons only rotation frequencies near the magnetron frequency are used, resulting in typical rotation frequencies of tens to hundreds of kilohertz \cite{Ball2019}. 
 
Doppler laser cooling and photon scattering for valence electron spin (qubit) state detection are provided by the same pair of lasers near $313\,\mathrm{nm}$, propagating parallel (axial) and perpendicular (radial) to the axis of crystal rotation, respectively. These lasers are typically red detuned by half the line width $\Gamma\approx19.6\,\mathrm{MHz}$ of the $2s^2S_{1/2}\xrightarrow{}2p^2P_{3/2}$ transition. The magnetic field $B_0$ produces an energy splitting of approximately $55\,\mathrm{GHz}$ between the valence electron spin states $\ket{2s^2S_{1/2},m_J=+1/2}=\ket{\uparrow}$ and $\ket{2s^2S_{1/2},m_J=-1/2}=\ket{\downarrow}$.  Millimeter waves at the $\ket{\uparrow} \leftrightarrow{} \ket{\downarrow}$ spin-flip transition frequency can be used to drive global spin oscillations. 

When an ion is in the $\ket{\uparrow}$ state, the Doppler cooling and detection lasers are near-resonant with the $\ket{\uparrow}\xrightarrow{}\ket{2p^2P_{3/2},m_J=+3/2}$ transition and scatter photons. This spin state is termed ``bright''. In contrast, when an ion is in the $\ket{\downarrow}$ state, the lasers are far off-resonance and produce almost no fluorescence photons. This spin state is termed ``dark''; ion imaging is based on detecting this scattered fluorescence. 

\section{ION IMAGING AND POSITION DETECTION} 
\label{sec.ion_detection}

To perform site-resolved quantum-state detection, individual ions must first be resolved, and their positions must then be localized. Our approach combines these two processes in an automated and robust way to support a high-data-rate experimental acquisition. 

The core efforts in this paper focus on top-view imaging parallel to the trap and crystal axis of rotation, through which individual ions in a 2D configuration may be spatially resolved [see Fig. \ref{fig:trap}] (side-view imaging via an electron-multiplying charge coupled device camera is used incidentally to analyze the crystal conformation). Ultraviolet (UV) photons at wavelength $313\,\mathrm{nm}$ are converted to visible (VIS) photons by an image intensifier (PP2360AL with Hi-QE Blue photocathode, P47 phosphor (half-life approximately $70\,\mathrm{ns}$) and a spatial resolution of $36\,\mathrm{lp/mm}$, in CRICKET adapter, Photonis Scientific Detectors) with a quantum efficiency of about $27\%$, which are then recorded by a TPX3CAM (Amsterdam Scientific Instruments). The TPX3CAM is a timestamping photon detector with 256x256 pixels, $(55 \times 55) \,\mu\mathrm{m}^2$ pixel size, $1.56\,\mathrm{ns}$ timing resolution, and a maximum data rate of 80Mhits/s \cite{Nomerotski2019, Poikela2014}, which has already been applied for fast detection in linear ion strings \cite{Zhukas2021, Zhukas2021a}. This imaging apparatus is placed outside the fringe fields of the superconducting magnet~\cite{Ball2019}, leveraging an objective lens system with a magnification of about $23$ and collection efficiency of $3.6\%$, which produces an image about $1.45\,\mathrm{m}$ away from the ion crystal. Ultimately, the resolution of the imaging system is about $2.4\,\mathrm{\mu m}$, limited by the pixel size and magnification. However, this is small compared to an ion-ion spacing of about $35\,\mathrm{\mu m}$ as shown in Fig. \ref{fig:trap}.  

The adjustable conversion gain for UV$\xrightarrow{}$VIS photons of the image intensifier is chosen as a compromise between detection efficiency, resolution, and total data rate. Every converted UV photon can create a large number of VIS photons to be recorded by the TPX3CAM. Imaging of a large number of ions simultaneously requires a UV$\xrightarrow{}$VIS photon conversion rate balancing the process of potentially missing detection of incident UV photons and saturating the VIS detector. 

\begin{figure}[h!tb]
    \centering
    \includegraphics[width=0.45\textwidth]{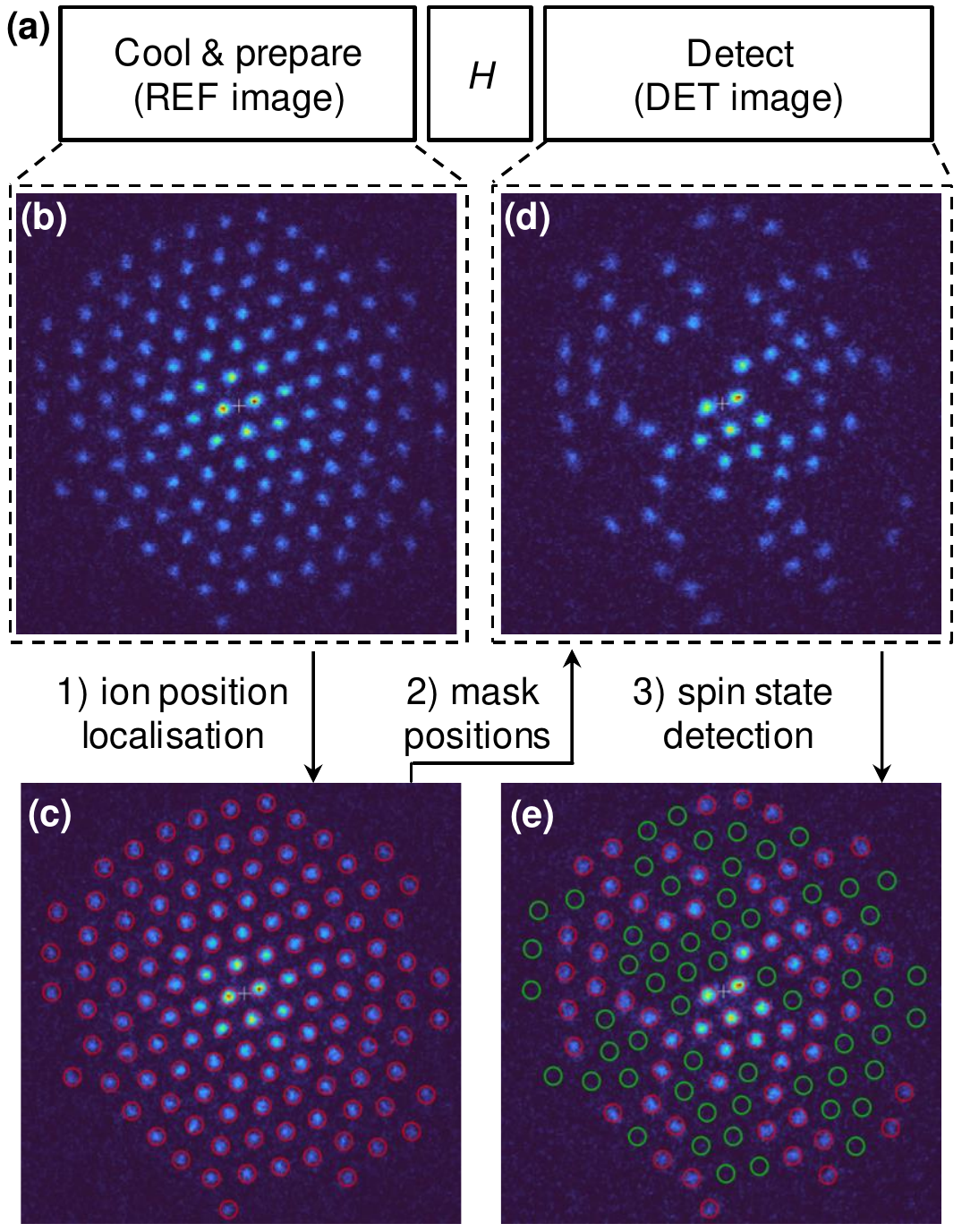}
    \caption{ (a) Generalized measurement sequence for single-shot ion position localization and spin-state detection. (b) During the initial Doppler laser cooling sequence, all ions are ideally in the bright state. The TPX3CAM records photon data from which a reference image (REF) is created through derotation to $t_0$. (c) A Faster R-CNN algorithm is used to localize the ion positions (step 1) which are stored for use in the final detection step. The image contains about 120 ions and is recorded for $20\,\mathrm{ms}$. A total of about 109\,000 VIS photons are detected, resulting in an average TPX3CAM data rate of $5.45\,\mathrm{Mhits/s}$ during detection, about one order of magnitude below the data-rate limit. Using DBSCAN, about 80\,000 UV photons were reconstructed. The innermost ions scatter about 40\,000 UV photons/s while the outermost about 10\,000 UV photons/s. (d) After the experimental interaction sequence is finished, the detection lasers are used for a projective spin-state measurement, resulting in a mixture of bright and dark ions. The recorded data are derotated to the same time $t_0$ as in the initial image, which would reproduce the crystal in the same rotation phase (DET image). Ideally, the ion positions have stayed the same during these steps. The previously found ion positions are masked to this image (step 2). (e) The time data in each ion region of interest can be evaluated to determine the spin state (step 3). This results in bright ions (spin state $\uparrow$) marked in red, while ions identified as dark (spin state $\downarrow$) are marked in green.}
    \label{fig:detection} 
\end{figure}

The TPX3CAM high-speed imaging system records the pixel position ($x_i,y_i$), time of arrival $t_i$ and time over threshold $t_{tot,i}$ of every detected photon $i$.  To resolve individual ions in a rotating crystal, a derotation coordinate transformation of the photon events is performed. This transformation employs the stable reference-crystal rotation frequency $\omega_r/2\pi$ set by the rotating wall drive and assumes stable relative ion positions and a known center of rotation ($x_c ,y_c$). The derotated photon coordinates are given by

\begin{equation}
    \begin{pmatrix}
    x_{R,i} \\
    y_{R,i}
    \end{pmatrix}
    =
    \begin{pmatrix}
    \cos{\alpha_i} & -\sin{\alpha_i} \\
    \sin{\alpha_i} &  \cos{\alpha_i} \\
    \end{pmatrix}
    \cdot
    \begin{pmatrix}
    x_i - x_c \\
    y_i - y_c
    \end{pmatrix},    
\end{equation}
where $\alpha_i=\omega_{r,\mathrm{TPX}} (t_i-t_0)$, with $t_0$ being an experiment-specific time offset and $\omega_{r,\mathrm{TPX}}=c_\mathrm{TPX}\omega_r$ being the crystal rotation frequency in the timebase of the TPX3CAM.  The rotation center of the ion crystal ($x_c, y_c$) is determined by employing a Nelder-Mead optimization routine which minimizes the inverse of the sum of a 2D gradient image filter (Sobel filter) as the cost function. 

The TPX3CAM used in this study does not accept an external clock signal, leading to unavoidable differences between the camera acquisition timing and the timebase of the rotating wall drive. We correct this by timestamping the rotating wall sync signal rising and falling edges via a dedicated time-to-digital converter input and adjusting the rotation frequency in the timebase of the TPX3CAM by $c_\mathrm{TPX}$. We find empirically that the relative frequency difference between the experimental master clock and the internal TPX3CAM clock is about 4 parts per million. 

The image intensifier creates a distribution of VIS photons for every detected UV photon. As such, some UV photons generate multiple VIS events closely linked in position and time at the TPX3CAM. We detect these photon-event clusters using a density-based spatial clustering of applications with noise (DBSCAN) algorithm with an arrival time difference scaling of $50\,\mathrm{ns}$ and a maximal Euclidean distance of 2 between two samples in a cluster. Photon clusters are used to characterize the so-called time-walk effect, which originates from constant-threshold discrimination to determine arrival times of signals with varying intensity. Although the signal intensity is not directly recorded, it is related to the recorded time over threshold $t_{tot,i}$. By analyzing the relation between $t_{tot,i}$ and $t_i$ over many photon clusters, the arrival times can be corrected \cite{Turecek2016}. We apply this correction to all arrival times $t_i$ depending on their corresponding $t_{tot,i}$. For further processing, the $t_{tot,i}$-weighted mean of the cluster's position, resulting in subpixel position resolution, is combined with the minimal arrival time among all photons in the cluster.

Single-shot photon data is binned to form a 256x256 pixel image, intensity-normalized, and analyzed by a ``Faster R-CNN'' algorithm~\cite{Ren2015} to identify the ion positions. Training data for the artificial neural network are generated using a range of ion crystal sizes and equilibrium-position configurations as would be realized with different trapping parameters~\cite{Wang2013}.  

These calculated position data are then overlapped with the measured fluorescence intensity distribution and background noise from scattered light found in our setup to produce faithful simulations of actual images from this experimental setup. This training approach allows the neural network to identify likely ion positions over a different number of ions, different underlying crystal structures, and radial nonuniformities in ion spacing. The training results in object detection metrics precision (correctly found ions divided by the number of all detected ions) and recall (correctly found ions divided by the total number of ions) of about 90\%.

We use the trained model to perform inference on experimental data to find the position of ions prepared in the bright state. An example of a typical experimental sequence is shown in Fig. \ref{fig:detection}. The initial Doppler cooling and preparation step produces a reference (REF) image [see Fig. \ref{fig:detection}a]. Ideally, all $^9\mathrm{Be}^+$ ions are maintained in the bright state via a dedicated repump laser which counteracts bright-state decay during the readout. The resulting derotated image [Fig. \ref{fig:detection}b] is then used for position inference, as can be seen in Fig. \ref{fig:detection}c, where the derotated image is annotated with artificial-neural-network-identified ion locations. The single-shot image in Fig.\ref{fig:detection}c contains about 120 ions and is recorded for $20\,\mathrm{ms}$. Large crystal conformation changes can be identified by an equilibrium-position-overlap discrepancy of bright ions in a reference and detection image pair and can be easily excluded from the data analysis. These may originate from, for example, crystal heating events or the formation of beryllium hydride~\cite{Sawyer2015} or other molecular ion species, both triggered by background gas collisions. The latter is typically followed by a mass-dependent centrifugal separation which causes a crystal reconfiguration.

Occasionally, we observe small variations between the crystal rotation phase and the phase of the rotating wall. This could be connected to slip-sick dynamics resulting from stress on the ion crystal due to the radial Doppler cooling laser and thermal effects \cite{Mitchell2001}. The resulting changes of ion positions between subsequent images are well below the width of single-ion-fluorescence spatial spread and can therefore be accounted for in the spin-state detection.   

\begin{figure}
    \centering
    \includegraphics[width=0.45\textwidth]{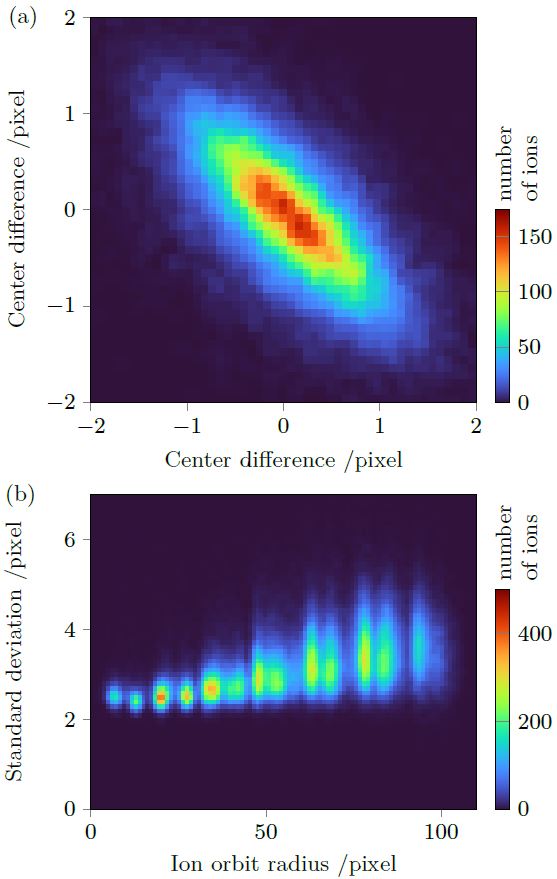}

    \caption{(a) Two-dimensional histogram of the horizontal and vertical distance between each ion position found by the neural network and its corresponding estimated position with a 2D Gauss distribution. The mean deviation in both directions is less than 1 pixel. (b) Two-dimensional histogram of standard deviations $\sigma_{1,2}$ of the major and minor axis of 2D Gauss distributions, fitted separately to each localized ion position, as a function of radial distance to the rotation center of the ion crystal. About 1200 images with more than 142\,000 ions have been evaluated in both data sets.}
    \label{fig:Fig_Gauss_combined} 
\end{figure}

Next, the ion positions are projected onto the measurement image (DET), taken after some arbitrary interaction $H$ [see Fig. \ref{fig:detection}d]. This image is derotated to the same time $t_0$ as the REF image, resulting in an ion crystal in the same rotation phase [see Fig. \ref{fig:detection}d]. To evaluate the accuracy of the individual ion positions found by the artificial neural network, $(X_{\mathrm{NN},i}, Y_{\mathrm{NN},i})$, we compare them to ion positions determined through least-squares fitting of 2D Gaussian distributions for every individual ion, using the artificial neural network positions as starting parameters for the fits. This results in the center values, $(X_{\mathrm{G},i}, Y_{\mathrm{G},i})$, as well as standard deviations along the major and minor axis, $(\sigma_{i,1}, \sigma_{i,2})$. The center differences in both spatial dimensions, $\Delta X_i = X_{\mathrm{NN},i} - X_{\mathrm{G},i}$ and $\Delta Y_i = Y_{\mathrm{NN},i} - Y_{\mathrm{G},i}$ are shown in Fig.~\ref{fig:Fig_Gauss_combined}a as a 2D histogram. In this comparison, the mean deviation between the identified ion positions derived from the two methods is less than 1 pixel, more than an order of magnitude smaller than the average ion-ion spacing of around $15$ pixels. A clear correlation between vertical and horizontal deviations in the center location is not yet explained, but this does not materially change the quality of the ion-position-identification procedure.

Calculating the standard deviations as above and plotting them as a histogram, we can observe the impact of ion radius on positioning accuracy (Fig. \ref{fig:Fig_Gauss_combined}b). This information is essential to inform each ion's region of interest employed in the spin-state analysis. The innermost ions have nearly equal standard deviations along both axes, $\sigma_{1,2}\approx 2.5\,\mathrm{pixels}$, while the outermost ions' area increases and becomes more elliptical, $\sigma_{1}\neq\sigma_{2}$. The ellipse's major axis is aligned with the tangential direction of motion, as seen directly in Fig.~\ref{fig:detection}d. We identify possible causes in the residual uncertainty in the time-walk correction or possible excitation of in-plane modes of the ion crystal~\cite{Wang2013}.


\begin{figure}%
    \begin{tikzpicture}%
        \begin{axis}[%
            name=MyAxis,
            xlabel={Detection time /ms}, ylabel={Detection fidelity},
            xmin=0, xmax=26, ymin=0.7, ymax=1,
            major tick length=\MajorTickLength,
            legend pos=south east, legend style={draw=none, fill=none, legend cell align=left},
            xtick pos=bottom, ytick pos=left,
            xtick align=outside, ytick align=outside,
            width=8cm, height=6cm
            ]
            \addplot+[black, only marks, mark=triangle, mark size=2, error bars/.cd, y dir=both, y explicit, error mark=-]%
            table[x expr=\thisrow{x}, y=fid_D, y error=fid_D_error] 
            {fidelities_error.txt};
            \addplot+[blue, only marks, mark=o, mark size=2, error bars/.cd, y dir=both, y explicit, error mark=-]%
            table[x expr=\thisrow{x}, y=fid_ave, y error=fid_ave_error] 
            {fidelities_error.txt};
            \addplot+[red, only marks, mark=square, mark size=2, error bars/.cd, y dir=both, y explicit, error mark=-]%
            table[x expr=\thisrow{x}, y=fid_B, y error=fid_B_error] 
            {fidelities_error.txt};
            \legend{dark $\mathcal{F}_D$, average $\mathcal{F}$, bright $\mathcal{F}_B$}
        \end{axis}%
        \node[right] at (MyAxis.left of north west) {(a)};
        \begin{axis}[%
            name=MyAxis2,
            xlabel={Detection time /ms}, ylabel={Counts},
            xmin=0, xmax=25, ymin=0, ymax=29,
            major tick length=\MajorTickLength,
            xtick pos=bottom, ytick pos=left, xtick align=outside, ytick align=outside,
            legend pos=north east, legend columns={4}, 
            legend style={draw=none, fill=none, legend cell align=right},
            width=8cm, height=4cm, at={($(MyAxis.south)-(0,1.3cm)$)}, anchor=north,
            ]
            \addplot+[const plot, red, fill opacity=0.1, fill=red, no marks]
            table[x expr=\thisrow{time}*1000, y=counts] {220824_115540_13_TPX_shots_mmW_tpx_RGB_1_40_ion85.txt}\closedcycle;
            \addplot+[const plot, black, fill opacity=0.1, fill=black, no marks]
            table[x expr=\thisrow{time}*1000, y=counts] {220824_115540_21_TPX_shots_mmW_tpx_RGB_1_75_ion40.txt}\closedcycle;
            \legend{bright, dark}
        \end{axis}%
        \node[right] at (MyAxis2.left of north west) {(b)};
    \end{tikzpicture}%

    \begin{tikzpicture}
        \begin{axis}[
            name=leftplot,
            xlabel={Ion orbit radius /pixel},
            ylabel={Detection time /ms},
            xmin=0, xmax=95,
            ymin=0, ymax=10,
            xtick={0,40,80},
            major tick length=\MajorTickLength,
            xtick pos=bottom, ytick pos=left,
            xtick align=outside, ytick align=outside,
            colormap/thermal, 
            view={0}{90},
            colorbar, colorbar horizontal,
            colorbar style={at={(0.5,1.03)}, anchor=south, xticklabel pos=upper,xtick={0.8,0.9}, height=0.2cm, },
            width=4cm, height=4cm,
            ]
            \addplot3 [surf,faceted color=none,]%
            table[x=x, y=y, z=z] {ML_time_binned_data_110_4_26_detection_bi_time_ave.txt};
        \end{axis}
        \node[left] at ($(leftplot.above north east)+(0.1cm,.17cm)$) {$\mathcal{F}$};
        \node[right] at ($(leftplot.left of north west)+(0,0.5cm)$) {(c)};
        \begin{axis}[
            name=rightplot,
            xlabel={ROI radius /pixel},
            ylabel={Time bins per 25ms},
            xmin=0.5, xmax=6.5,
            ymin=2.5, ymax=10.5,
            major tick length=\MajorTickLength,
            xtick pos=bottom, ytick pos=left,
            xtick align=outside, ytick align=outside,
            colormap/thermal, 
            view={0}{90},
            colorbar, colorbar horizontal,
            colorbar style={at={(0.5,1.03)}, anchor=south, xticklabel pos=upper,xtick={0.92,0.94}, height=0.2cm, },
            width=4cm, height=4cm,
            at={($(leftplot.outer east)+(0cm,0cm)$)},anchor=outer west,
            ]
            \addplot3 [matrix plot*]
            table[x=x, y=y, z=z] {ML_error_results_backup_ROI_ave.txt};
        \end{axis}
        \node[left] at ($(rightplot.above north east)+(0.1cm,.15cm)$) {$\mathcal{F}$};
        \node[right] at ($(rightplot.left of north west)+(0,0.5cm)$) {(d)};
    \end{tikzpicture}
    \caption{Spin-state detection in a crystal of 120 ions. (a) Spin-state detection fidelity as a function of detection time using time bins of $T_B=1\,\mathrm{ms}$ and a region of interest radius of $r_\mathrm{ROI}=4\,\mathrm{pixels}$. The error bars represent the standard deviation over all ion orbit positions. (b) Time-binned photon counts of an ion detected bright and an ion detected dark which are optically pumped into the bright state by the detection lasers after about $9\,\mathrm{ms}$. (c) Average spin-state detection fidelity (color) as a function of ion orbit radius and detection time using time bins of $T_B=1\,\mathrm{ms}$ and $r_\mathrm{ROI}=4\,\mathrm{pixels}$. (d) Average spin-state detection fidelity (color) for different region-of-interest radii $r_\mathrm{ROI}$ and number of time bins per $25\,\mathrm{ms}$ detection time. See text for further discussion.}
    \label{fig:Fig_state_detect}
\end{figure}
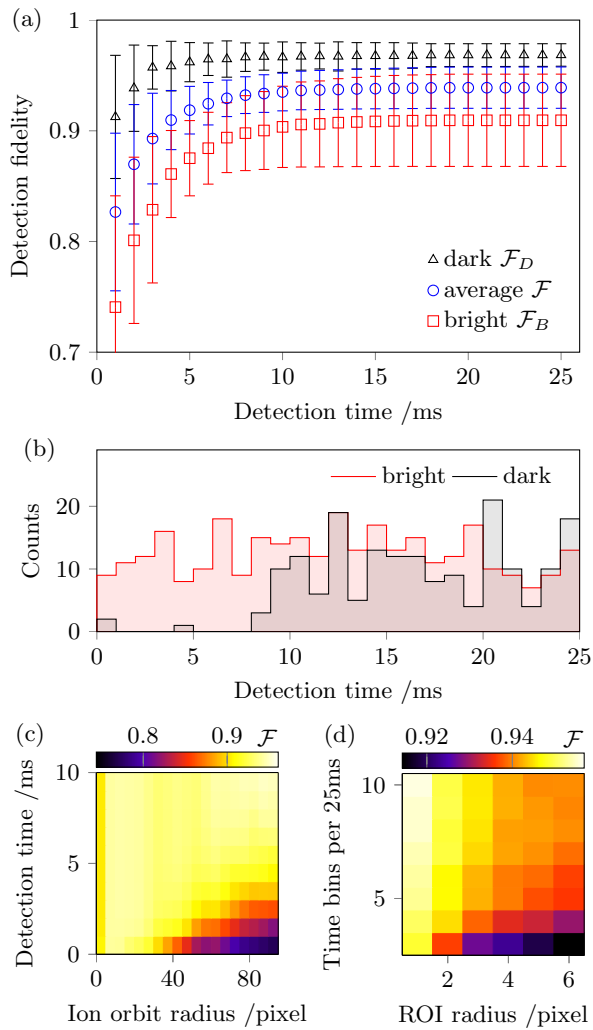%

\section{SPIN-STATE DETECTION} 

Ion-spin-state determination relies on the discrimination of dark and bright fluorescing ions. In order to allow direct state detection for ions that are not fluorescing and hence cannot be directly localized, we compare all images to a reference image taken a  few milliseconds before each detection sequence. This approach mitigates the effects of slow drifts or dynamic restructuring in the ion crystal while not contributing significantly to the length of an experimental iteration.

State discrimination employs a time-resolved approach to account for fluorescence dynamics occurring during the readout procedure. In particular, state leakage during measurement is mitigated using the time-resolved maximum likelihood method described by W\"olk \textit{et al.}~\cite{Woelk2015}. To this end, the bright state photon rate $R_B$, the dark state photon rate $R_D$, the bright state lifetime $\tau_B$, and the dark state lifetime $\tau_D$ must be determined during an initial calibration procedure. These parameters depend on the detection lasers' spatial, temporal, and spectral overlap with the respective ion. For example, the Doppler-shift-induced nonuniformity of the fluorescence-photon distribution from the radial detection laser produces fluorescence in the form of a narrow stripe in the laboratory frame of reference (see Fig. \ref{fig:trap}), which becomes a radially varying distribution in a corotating frame. 

To determine the radius-dependent bright- and dark-state scattering rates $R_B(r)$ and $R_D(r)$ respectively, we prepare ion crystals with about 120 ions and a diameter of around $480\,\mathrm{\mu m}$ (approximately 200 pixels) first in the bright state and then in the dark state via application of a millimeter-wave $\pi$-pulse. Global fluorescence detection estimates state-preparation fidelity at $>99\%$. After ion localization in the bright-state image, we define a detection region-of-interest (ROI) of radius $r_\mathrm{ROI}$ around every ion and divide the captured photon events according to their arrival time $t$ in $N$ time bins of duration $T_B$. The resulting count rate in the first time bin of every ion is plotted against the ion orbit radius $r_i=\sqrt{X_{\mathrm{NN},i}^2 + Y_{\mathrm{NN},i}^2}$; this protocol is performed for hundreds of images, and finally a polynomial fit to the data allows us to extract $R_B(r) \rvert_{r_\mathrm{ROI},T_B}$ for a specific combination of $(r_\mathrm{ROI},T_B)$. The same procedure is repeated after dark-state preparation to determine the dark-state scattering rate $R_D(r) \rvert_{r_\mathrm{ROI},T_B}$.

The radius-dependent lifetimes $\tau_B(r)$ and $\tau_D(r)$ are determined with the same datasets used to determine the scattering rates. First, we bin the photon events according to their arrival time (as above) and, in addition, according to their radius to the crystal center. Then, for every radius bin in the bright and dark ion image, we perform a simultaneous fit of an exponential decrease with offset to the bright photon events and an exponential increase with offset to the dark photon events over all time bins \cite{Woelk2015}. This allows extraction of both state lifetimes for every radius bin. Finally, we fit a polynomial function to the lifetimes for different radii, resulting in $\tau_B(r)$ and $\tau_D(r)$.  

We determine spin-state-detection fidelity by preparing pairs of ion images following bright-state and dark-state initialization. The bright-state image is used to determine the ion positions and the total ion number $N$. The time-binned method, as described above, is used to determine the number of bright (dark) ions $N_B$ ($N_D$) at the beginning of the detection sequence for each state preparation. Assuming negligible state preparation errors, we determine the bright (dark) state fidelity, $\mathcal{F}_B=N_B/N$ ($\mathcal{F}_D=N_D/N$). The average detection fidelity is given by $\mathcal{F}=(\mathcal{F}_B+\mathcal{F}_D)/2$. Figure \ref{fig:Fig_state_detect}a shows the bright, dark, and averaged detection fidelity as a function of the total detection time for time bins of $T_B=1\,\mathrm{ms}$ and a region of interest radius of $r_\mathrm{ROI}=4\,\mathrm{pixels}$. Figure \ref{fig:Fig_state_detect}b shows time-binned data of an ion initially in the bright and dark state. Due to optical pumping by the detection lasers, the state of the latter changes after about $9\,\mathrm{ms}$.

For an ion crystal size of $118\pm3$ ions, we calculate an average detection fidelity of $94(2)\%$  after a detection time of about $10\,\mathrm{ms}$. We observe substantial variability in detection fidelity with ion orbit radius, which can be compensated by increasing the detection time [Fig.~\ref{fig:Fig_state_detect}c];  beyond $10\,\mathrm{ms}$ we find less than $1\%$ variability in detection fidelity. The smallest radii statistics are very low due to a missing ion at the very center of the crystal. 

We emphasize that due to imperfections in the ion position detection, which result in the previously discussed precision and recall of $90\%$, not all ions can be incorporated in the state detection analysis since their position is unknown. As a result, we treat position detection and spin-state detection as separate metrics. 


Our analysis identifies a trade-off between ROI for state detection and the neural network in identifying ion positions. Figure \ref{fig:Fig_state_detect}d shows the average detection fidelity as a function of the region-of-interest radius $r_\mathrm{ROI}$ and the number of time bins. Ensuring using at least 10 time bins leads to average detection fidelities of $94\%$. Reducing the ROI leads to an increase in detection fidelity to approximately $96\%$, which we ascribe to a diminished influence of background counts on the readout. In order to keep the ROI well within the ion-localization deviation of the neural network and match the spot size found earlier (Fig. \ref{fig:Fig_Gauss_combined}), a conservative $r_\mathrm{ROI}=4\,\mathrm{pixels}$ is selected.

\begin{figure}[t]
\input{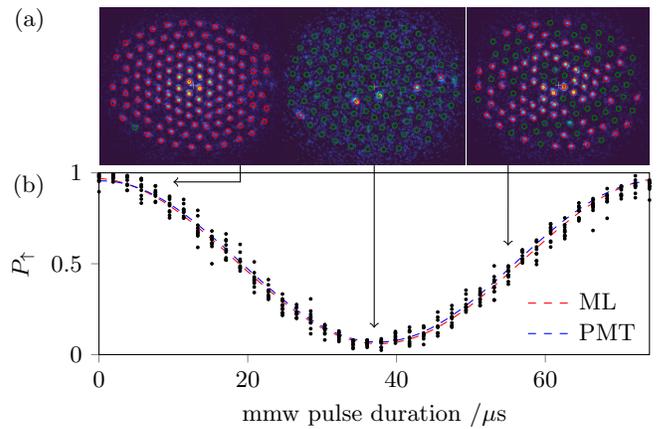}
\caption{Individual spin-state detection after global spin-flip excitations. (a) A crystal of about 120 ions is shown at different millimeter wave (mmw) pulse durations of $t=0$, $t\approx 39\,\mathrm{\mu s}$ and $t\approx 57\,\mathrm{\mu s}$, as indicated by the arrows. (b) Bright-state fraction, $P_\uparrow=N_B/N$, determined from individual images (dots) by counting the total number of ions, $N$, via the neural network, and the number of bright ions, $N_B$, via the time-binned maximum likelihood (ML) method. The red dashed line is a sinusoidal fit to the data. The blue dashed line is a fit to the photon counts (not shown) obtained simultaneously with a photomultiplier tube (PMT). The crystal's boundary has a higher ellipticity due to a stronger rotating wall potential, compared to Fig. \ref{fig:detection}.}
\label{fig.Rabi}
\end{figure}

Finally, we employ the detection scheme outlined above and shown in Fig. \ref{fig:detection} to detect individual spin states after different pulse durations of the $55\,\mathrm{GHz}$ millimeter-wave system tuned to the spin-flip transition frequency, thereby performing global Rabi oscillations. Figure \ref{fig.Rabi}a shows crystal images after three different pulse durations. Ions identified as bright ($\uparrow$ spin-state) are marked in red, while those identified as dark ($\downarrow$ spin-state) at the beginning of the detection time are marked in green. Figure \ref{fig.Rabi}b shows the fraction of the counted number of bright ions in each image, $P_\uparrow$. The fitted oscillation agrees well with the fitted photon counts obtained simultaneously with a photomultiplier tube. 

\section{SUMMARY AND OUTLOOK} 
This paper introduces an automated procedure for the readout and state discrimination in large crystals of trapped ions. The method employs recently developed hardware and advanced machine-learning techniques to identify ion positions in the extreme case of a Penning trap with kilohertz-level ion-crystal rotation. The method permits ion localization more than an order of magnitude below the ion spacing and delivers about $94\%$ state-readout fidelity. The techniques employed here apply to large ion crystals in Paul and Penning traps and neutral atom arrays in optical lattices.

One exciting opportunity pertains to improving the image-processing throughput via the choice of coding language and hardware for execution. In our experiments, all data processing is conducted in Python, where user-friendliness was prioritized before computational speed. Derotation of $10^5$ photons and the inference of ion positions on a desktop PC take about $1\,\mathrm{s}$ each and are therefore performed mainly in postprocessing. To spot-check the data quality, the control software displays one image roughly every $2\,\mathrm{s}$. We believe computation times can be significantly reduced by employing a GPU for inference and performing the derotation in a faster language.    

\section*{Acknowledgements}

We thank Dominic Jones, Finlay Dalton, William Dodds, Xanda Kolesnikow, and Yiwen Yang for their early work on the Faster R-CNN ion position detection and Michael Robinson for implementing the TPX3CAM. Furthermore, we thank Christophe Valahu for his help in setting up the Faster R-CNN training at the University of Sydney Artemis HPC. The project was supported by the Australian Research Council Centre of Excellence for Engineered Quantum Systems (CE170100009) and a private
grant from H. and A. Harley. R.N.W. acknowledges support from the Australian Research Council under the Discovery Early Career Researcher Award scheme (DE190101137). J.H.P and J.Y.Z.J. acknowledge support through the Sydney Quantum Academy. This material is based upon work supported by the Air Force Office of Scientific Research under award number FA2386-23-1-4067.

\bibliography{references}


\end{document}